\documentclass[prb,twocolumn,showpacs,superscriptaddress]{revtex4}
\usepackage{graphicx}
\usepackage{subfigure}

\begin{document}

\title{New Numerical Results Indicate a Half-Filling SU(4) Kondo State in Carbon Nanotubes}

\author{C. A. B\"usser}
\affiliation{Condensed Matter Sciences Division, Oak Ridge National Laboratory, Oak Ridge,
Tennessee 37831}
\affiliation{Department of Physics and Astronomy, The University of Tennessee, Knoxville,
Tennessee 37996}
\author{G. B. Martins}
\email[corresponding author: ]{martins@oakland.edu}
\affiliation{Department of Physics, Oakland University, Rochester, MI 48309}

\begin{abstract}
Numerical calculations simulate transport experiments in carbon nanotube 
quantum dots (P. Jarillo-Herrero {\it et. al}, Nature {\bf 434}, 484 (2005)), where 
a strongly enhanced Kondo temperature ${\rm T_K} \approx 8.0 ~{\rm K}$ was associated with the 
SU(4) symmetry of the Hamiltonian at quarter-filling 
for an orbitally double-degenerate single-occupied electronic shell. 
Our results clearly suggest that the Kondo conductance measured for an adjacent shell 
with ${\rm T_K} \approx 16.0 ~{\rm K}$, interpreted as a singlet-triplet 
Kondo effect, can be associated instead to an SU(4) Kondo effect at {\it half-filling}. Besides presenting 
spin-charge Kondo screening similar to the quarter-filling SU(4), the half-filling SU(4) has been 
recently associated to very rich physical behavior, including 
a non-Fermi-liquid state (M. R. Galpin {\it et al.}, Phys. Rev. Lett. {\bf 94}, 186406 (2005)). 
\end{abstract}

\maketitle

\section{Introduction}

The synthesis of nanostructures such as quantum dots (QDs) has attained a 
high level of sophistication, allowing control over systems displaying complex many-body 
properties. Recently, the observation of the Kondo effect in 
orbitally degenerate carbon nanotube (CNT) QDs by Jarillo-Herrero {\it et al.} 
\cite{jarillo} has renewed interest in the so-called 
SU(4) Kondo effect. 
Early measurements of orbital Kondo effect in double QDs can be found
in work by U. Wilhelm {\it et al.} \cite{wilhelm}, while more recent results
are reported in A. W. Holleitner {\it et al.} \cite{holleitner}. 
However, no conclusive evidence for an SU(4) Kondo effect in double QDs has been
established. To date, besides the results in CNT QDs \cite{jarillo}, clear evidence
of SU(4) Kondo has been reported in vertical QDs \cite{sasaki1}. 
Early theoretical work can be found in T. Pohjola {\it et al.} \cite{pohjola} and
in L. Borda {\it et al.} \cite{borda}, while a 
review of SU(4) Kondo in nanostructures was written by G. Zarand \cite{zarand}.
Recently, a flurry of theoretical results exploring more detailed
aspects of the SU(4) Kondo effect have been presented for a diversity of setups \cite{lehur}.
Quite recent transport measurements in ambipolar semiconducting CNT QDs \cite{gleb} 
report conductance results for a large sequence of electronic shells in the QD, with 
clear indication of SU(4) states.  

Besides the fact that the Kondo temperature of 
an SU(4) Kondo state is in general at least one order of magnitude higher than the 
traditional SU(2) Kondo temperatures \cite{note1}, there is also great interest in 
studying mesoscopic 
systems with two or more interacting SU(4) Kondo impurities, since this could 
shed light into the puzzling behavior of some bulk systems displaying the orbitally 
degenerate Kondo effect. For instance, $\rm Ce_x La_{1-x} B_6$, a well-known Kondo system 
with orbitally degenerate impurities, presents a magnetic phase diagram which 
still defies theoretical description \cite{iroi}.
Another intriguing aspect recently discussed is the possibility of 
orbitally degenerate QDs being Jahn-Teller active \cite{toonen}. In addition, the simultaneous 
Kondo screening of charge and spin, resulting in a many-body entangled state 
for these two degrees of freedom \cite{goldhaber}, points to the exciting 
possibility of observing new many-body states.

The interpretation of the CNT experimental results \cite{jarillo} through numerical 
calculations has concentrated specifically 
on quarter-filling (QF) (1 electron occupying the topmost electronic shell in the QD) 
\cite{aguado1, aguado2}. In reality, most of the theoretical research on the SU(4) 
Kondo effect in QDs has concentrated on the QF regime, with the sole exception of 
the work by M. R. Galpin {\it et al.} \cite{galpin1, galpin2}, where NRG 
calculations analyzed the properties of the SU(4) Kondo effect at half-filling (HF), i.e., 
with 2 electrons in the topmost electronic shell. 

In this paper, motivated by these interesting new possibilities regarding the 
SU(4) Kondo effect in QDs, the authors will use a recently developed numerical
method, called Embedded Cluster Approximation (ECA) \cite{method}, to reanalyze 
the conductance measurements performed in a CNT QD by 
Jarillo-Herrero {\it et al.} \cite{jarillo} and also to extend 
the already mentioned QF \cite{aguado1,aguado2} and HF \cite{galpin1,galpin2} 
results to all fillings, paying special attention to 
the robustness of the SU(4) state in respect to the tunneling properties 
of the QD. 
The rest of the paper is organized as follows: The model used is presented
in section II, where the ECA method will be briefly described.
As an illustration of the capabilities of the ECA method, in section III the
authors qualitatively reproduce the NRG results presented by Galpin {\it et al.}
\cite{galpin1}.  
In section IV it will be shown that the state where the orbital degree
of freedom is not conserved upon tunneling (see Fig. 1), also called Two-Level SU(2) (2LSU(2)),
is qualitatively different from the SU(4) state (where the orbital degree of freedom
is conserved) even at zero magnetic field. Also in section IV, 
the authors will analyze the transition between SU(4) and 2LSU(2) states, 
comparing our results to previously published results \cite{aguado2}.
In section V, by realizing that there {\it is} spin-charge entanglement also at
HF, the authors will suggest that the conductance of one of the electronic
shells observed in the experiments (the third shell) can be associated to an SU(4) 
Kondo effect, offering an alternative to the single-triplet effect \cite{sasaki} interpretation
suggested by Jarillo-Herrero {\it et al.} \cite{jarillo}.
The authors also note that recent experimental results by Makarovski {\it et al.}
\cite{gleb} n CNT QDs give support to our HF SU(4) Kondo interpretation.
This new interpretation of the CNT results \cite{jarillo} implies that the rich 
physics unveiled by the NRG results
of Galpin {\it et al.} \cite{galpin1}, also confirmed by our results (see Fig. 2), could,
at least in principle, be probed in CNT QDs.
To further support our interpretation, in section V results in agreement with the experiments will
be presented for a magnetic field applied along the axis of the CNT \cite{jarillo,jarillo2}.
In section VI the conclusions are presented.

\section{Model}

\begin{figure}[h]
\centering
\includegraphics[height=4.0cm]{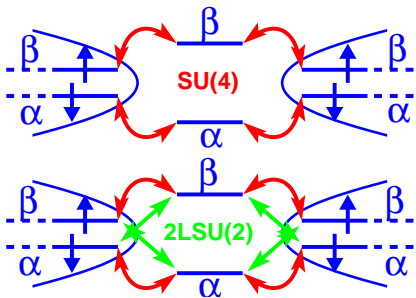}
\caption{ Schematic representation of the system being analyzed. Top: 
Hopping matrix elements (in red) conserving the orbital degree of freedom 
(indicated by the blue arrows) lead to a Hamiltonian with SU(4) symmetry.
Bottom: The orbital degree of freedom is not conserved 
upon tunneling (notice the green arrows), leading to a so-called 2LSU(2) state 
(see reference \onlinecite{aguado2} for a detailed discussion of this 
transition at QF).
}
\end{figure}

The CNT QD will be modeled by an orbitally degenerate Anderson impurity 
coupled to leads with two conduction channels:
\begin{eqnarray}
H_{\rm d}=\sum_{\sigma;\lambda=\alpha,\beta}  
\left[ {U \over 2} n_{\lambda \sigma} n_{\lambda \bar{\sigma}} + V_g n_{\lambda \sigma}\right]
+ U^{\prime} \sum_{\sigma \sigma^{\prime}} n_{\alpha \sigma} n_{\beta \sigma'} ,
\end{eqnarray}
\begin{eqnarray}
H_{\rm leads} &=& t \sum_{l=R,L} \sum_{\sigma;\lambda=\alpha,\beta;i} 
\left[ c_{l_{\lambda} i\sigma}^{\dagger} 
c_{l_{\lambda} i+1\sigma} +\mbox{h.c.} \right], \\
H_{\rm int} &=& \sum_{l=R,L} \sum_{\lambda;\lambda^{\prime};\sigma} t_{\lambda \lambda^{\prime}}
\left[ d_{\lambda \sigma}^{\dagger} c_{l_{\lambda^{\prime}} 0\sigma} + \mbox{h.c.} \right] ,
\end{eqnarray}
where $H_{\rm d}$ describes the orbitally degenerate Anderson impurity, subjected to a gate potential 
$V_g$, and the second and third equations describe the leads and their interaction with the 
CNT QD, respectively. More specifically, $\lambda=\alpha, \beta$ are 
two degenerate orbitals associated to the wrapping mode (clockwise or counterclockwise) 
of the electron propagation along the axial direction of the CNT \cite{minot}, 
while $d_{\alpha \sigma}$ ($d_{\beta \sigma}$) annihilates an electron with spin $\sigma$ in the 
$\alpha$ ($\beta$) orbital in the CNT and $c_{l_{\alpha}i\sigma}$ ($c_{l_{\beta}i\sigma}$) 
annihilates an electron with spin $\sigma$ in the i-th site of the 
$\alpha$ ($\beta$) channel in the $l=R, L$ (right or left) lead \cite{channels}.
We introduce intra- and inter-orbital Coulomb repulsions $U$ and $U^{\prime}$, respectively. 
To decrease the number of parameters in the model, the hopping matrix elements connecting 
the CNT QD to the leads (eq. 3) are the same at left and right, 
and assumed to follow the equalities $t_{\alpha \alpha}=t_{\beta \beta}=t^{\prime}$ (red 
arrows in Fig. 1) 
and $t_{\alpha \beta}=t_{\beta \alpha}=t^{\prime \prime}$ (green arrows in Fig. 1). As discussed in 
references \onlinecite{aguado1} and \onlinecite{ aguado2}, 
when $t^{\prime}$ is finite and $t^{\prime \prime}=0$, 
one has an SU(4) Kondo state. On the other hand, when $t^{\prime}=t^{\prime \prime}$, 
one has the so-called 2LSU(2) Kondo state.
As discussed in the Supplementary Information in reference
\onlinecite{jarillo}, there are fundamental differences between these two states.
The 2LSU(2) Kondo effect is more akin to the singlet-triplet Kondo effect \cite{sasaki},
where spin and orbital degrees of freedom have different roles: only the spin degree of freedom
is screened by the conduction electrons, while the degenerate orbital levels just
contribute to the increase in the possible number of co-tunneling processes,
leading to an enhanced Kondo temperature. In this case, the orbital degree
of freedom is {\bf not} screened by the conduction electrons. On the other
hand, in the SU(4) state, spin and orbital degrees of freedom
participate in the same footing. Both of them are screened by the conduction
electrons, and if the screening of one of the degrees of freedom is somehow
suppressed, the system is then left in an SU(2) Kondo effect stemming from
the other degree of freedom.

To calculate the conductance $G$, using the Keldysh formalism \cite{Meir-cnd}, 
a cluster containing the orbitally degenerate Anderson impurity plus a few sites of 
the leads is solved exactly, the Green 
functions are calculated, and the leads are then incorporated through a Dyson Equation 
embedding procedure \cite{method}. All the results shown were obtained for 
(in units of $t$) $U=0.5$, $t^{\prime}=0.2$, 
zero-bias, and zero temperature. The value of $t^{\prime \prime}$ varies between zero and 
$t^{\prime}$ and most of the results shown are for $U^{\prime}=U$. 
When a magnetic field is applied along the CNT axis, 
besides the Zeeman splitting coming from the spin degree of freedom, the orbital 
levels behave as a pseudo-spin $1/2$ and are also split \cite{minot}:  
$H_{\rm Zeeman} = B\left[\mu_{orb} \sum_{\sigma} \left(n_{\beta\sigma}-n_{\alpha\sigma}\right)
+ \mu_{sp} \sum_{\lambda} \left(n_{\lambda+}-n_{\lambda-}\right)\right]$. 
As reported by Jarillo-Herrero {\it et al.} \cite{jarillo}, the 
orbital magnetic moment $\mu_{orb}$ experimentally measured is such that the orbital 
splitting is one order of magnitude larger than the spin splitting.
For the actual simulation of the experimental results (Figs. 6 and 7), the degeneracy 
of the orbital levels will be raised by introducing a small energy 
splitting $\delta E$ \cite{splitting}.

\section{Comparison with NRG results at Half-Filling}

Before presenting the main results in this work, the authors 
will show that ECA can qualitatively reproduce the NRG results of Galpin 
{\it et al.} (Fig. 2). Besides showing that ECA captures correctly the 
physics of the model, this qualitative agreement shows that the NRG 
low-energy results in ref. \onlinecite{galpin1} are robust, suggesting
that they could be observable if one can find an experimental realization 
of the SU(4) state at HF (more on that below).

\begin{figure}[h]
\centering
\includegraphics[angle=90,height=6.5cm]{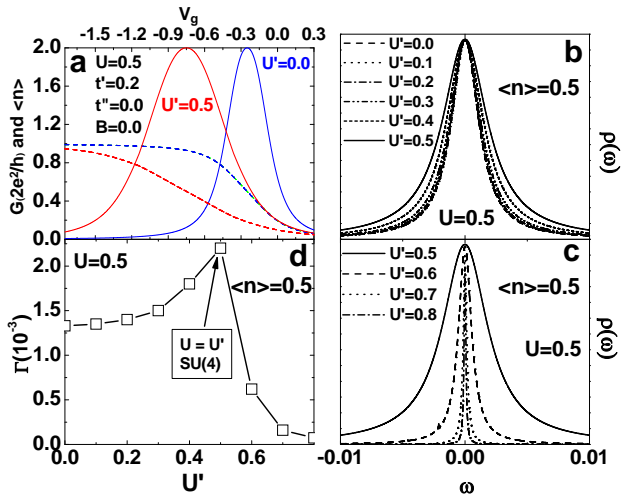}
\caption{  
(a) Blue: Conductance $G$ (solid) and average occupancy $\left< n \right>$ 
per spin orientation, per orbital (dashed) obtained when the two orbital levels are
not correlated ($U^{\prime}=0.0$). Red: Same as blue curves, but
now the orbital levels are correlated ($U^{\prime}=0.5$). In both cases,
$U=0.5$, $t^{\prime}=0.2$, and $t^{\prime \prime}=0.0$.
(b) Variation with $U^{\prime}$ of the LDOS at HF, from $U^{\prime}=0.0$ (SU(2)$\times$SU(2),
bottom curve) to $U^{\prime}=U=0.5$ (SU(4), top curve).
(c) Same as in (b), but now for $U^{\prime}\geq U$ ($U^{\prime}=U$ top curve).
(d)  Variation with $U^{\prime}$ of the width of the peaks ($\Gamma$) in (b) and (c), which is proportional
to $T_K$. The qualitative behavior matches the results presented in reference \onlinecite{galpin1}.
}
\end{figure}

Figure 2 displays the influence of $U^{\prime}$ (Coulomb repulsion 
between the two orbital levels) over the conductance in the case 
where the two channels are independent ($t^{\prime \prime}=0.0$), 
i. e., a transmitted electron that tunnels into the CNT QD coming from the 
$\alpha$ ($\beta$)-channel in the left lead, can only tunnel out through 
the $\alpha$ ($\beta$)-channel in the right lead. The blue curves 
($U^{\prime}=0$) in Fig. 2a are representative of two {\it independent} spin SU(2) Kondo 
effects (associated to each channel) which are simply added together. 
In this case, each channel magnetically screens the spin situated 
on the level to which it is connected ($\alpha$ or $\beta$). This 
situation changes if the orbital levels are 
correlated with each other (finite $U^{\prime}$, red curves), i. e., in the 
SU(4) state. As can be seen in Fig. 2b (at HF, i.e., $\left< n \right> =0.5$), 
the width of the Kondo resonance 
in the local density of states (LDOS) becomes larger as $U^{\prime}$ increases, indicating 
an enhancement of $T_K$: in the SU(4) state at HF, both 
degrees of freedom (spin and orbital) are participating in the Kondo effect 
and are being simultaneously screened (magnetically and electrostatically) by 
the conduction electrons \cite{galpin1}. Similarly to QF (as described in Fig. 1 
of Jarillo-Herrero {\it et al.} \cite{jarillo}), the increase in the number 
of possible co-tunneling processes now available 
for electron transport through the CNT QD results in an enhanced Kondo effect \cite{range}. 
Fig. 2c shows the abrupt suppression of the Kondo resonance for $U^{\prime} > U$. 
The difference between the 
regions above and below the SU(4) point is more clearly seen in Fig. 2d, where the width 
of the LDOS peaks in 2b and 2c (which is proportional to $T_K$) is plotted. 
This result is in qualitative agreement with ref. \onlinecite{galpin1}.

\begin{figure}[h]
\centering
\includegraphics[angle=90,height=3.5cm]{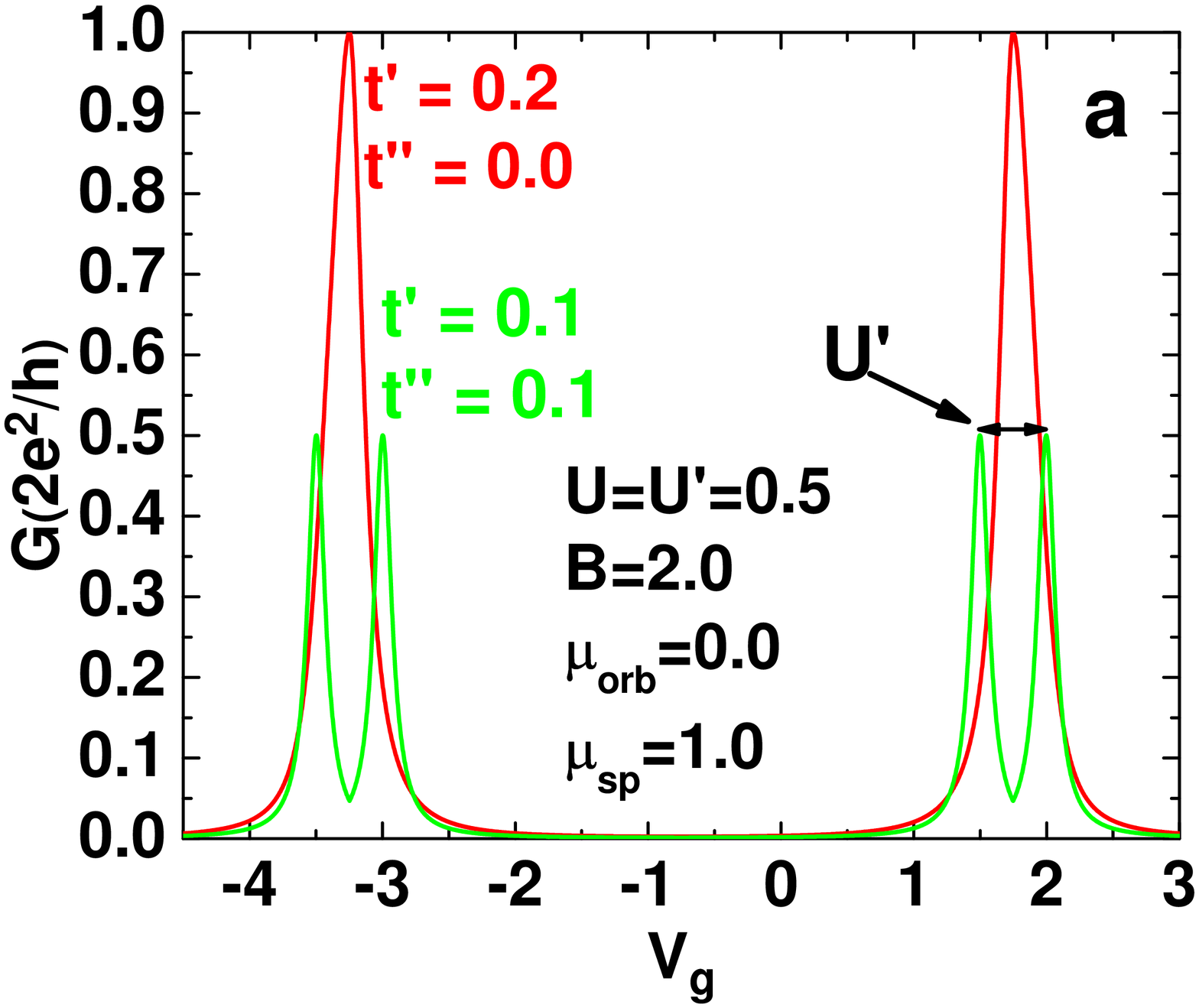}
\includegraphics[angle=90,height=3.5cm]{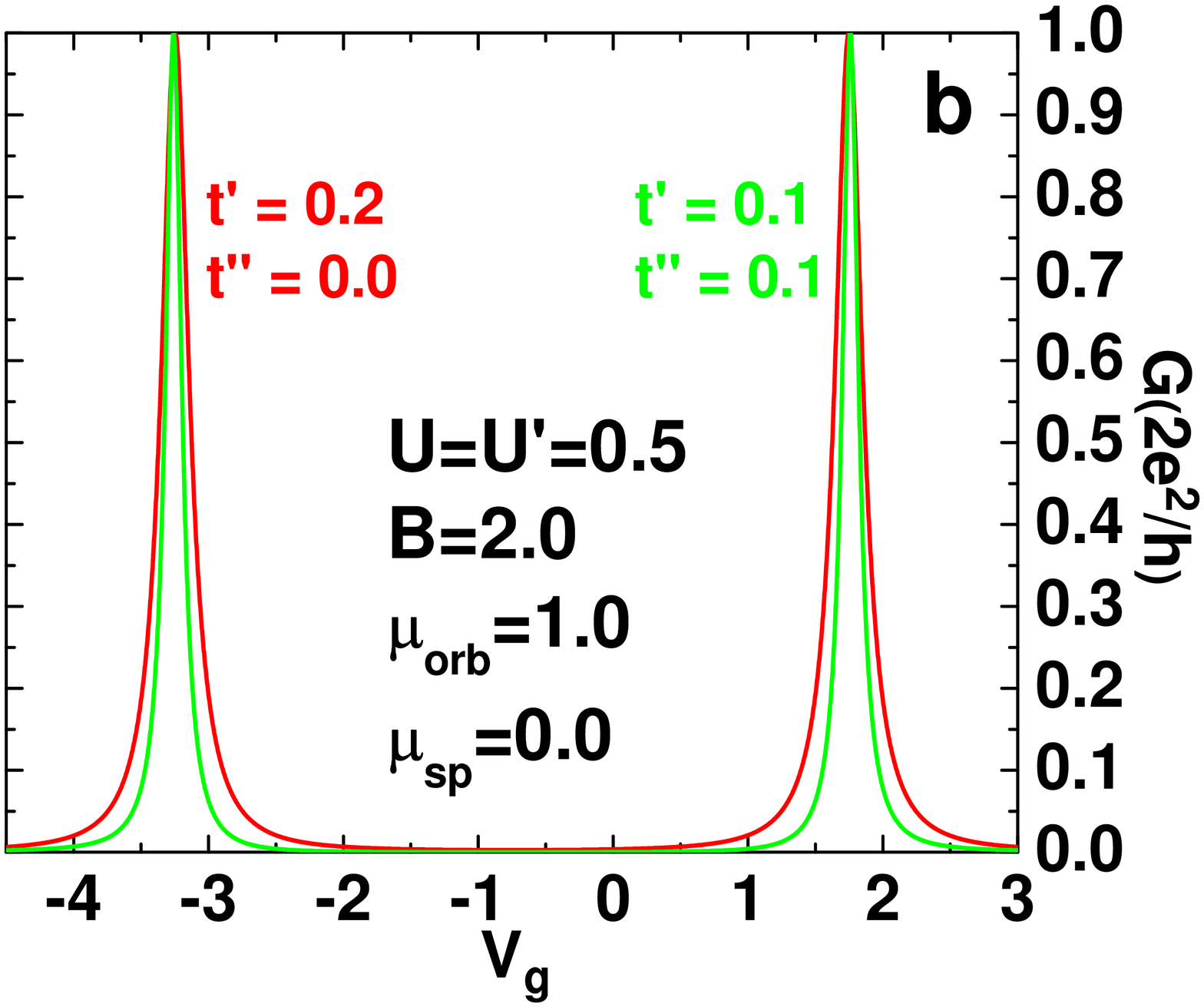}
\caption{ Illustration of the difference between SU(4) and
2LSU(2) Kondo states. (a) By applying an external magnetic field
and making $\mu_{orb}=0.0$ and $\mu_{sp}B=2.0$, one suppresses
the spin Kondo effect. In this case, the SU(4) Kondo peak seen
in Fig. 2 (red curve) splits into two orbital SU(2) Kondo peaks (red
curve), where the degree of freedom being screened is the 
orbital one. However, in the case of the 2LSU(2) (green curve), as the spin
Kondo effect has been suppressed by the field and since 
there is no orbital Kondo effect in the 2LSU(2) state, one
can see the two sets of CB peaks separated by $U^{\prime}$ (green curve).
(b) On the other hand, when $\mu_{orb}B=2.0$ and $\mu_{sp}=0.0$, one is left with
a spin SU(2) Kondo effect for each orbital level in both states (SU(4) and 2LSU(2)),
as is clearly shown by the two Kondo peaks separated by a distance proportional to the
applied field.
}
\end{figure}

\section{Comparison of SU(4) with 2LSU(2) and transition between both states}

To illustrate the entanglement of the charge and spin degrees 
of freedom in the SU(4) state, we compare results for $t^{\prime \prime}=0$ and 
$t^{\prime \prime}=t^{\prime}$, i.e., for the SU(4) and  2LSU(2) 
states (see Fig. 1), respectively. In Figs. 3a and 3b it is shown a comparison between them when
a magnetic field acts only on the spin degree of freedom (Fig. 3a) or
only on the orbital degree of freedom (3b). By making $\mu_{orb}=0.0$ and
$\mu_{sp}B=2.0$ (3a), the spin Kondo effect is suppressed, since the spin levels are
split by an energy larger than the Kondo temperature, while the orbital levels are
unaffected. Therefore, the SU(4) Kondo peak splits into two
{\it orbital} SU(2) Kondo peaks separated by a distance proportional to the field 
(red curve) \cite{lipinski}. 
On the other hand, since in the 2LSU(2) state there is no orbital
Kondo effect (since the orbital quantum number is not conserved upon tunneling),
by suppressing the spin Kondo with the magnetic field one is left with just
two sets of Coulomb Blockade (CB) peaks (split by $U^{\prime}$)
separated by a splitting proportional to the field (green curve). In contrast, when only the
orbital levels are split by the magnetic field ($\mu_{orb}B=2.0$ and $\mu_{sp}=0.0$),
one is left with two spin SU(2) Kondo peaks (one for each
orbital level) for both states, as can be seen in Fig. 3b.

At zero field, it is then not really surprising that the SU(4) state depicted 
by the red curve in Fig. 2a will change once the channels are allowed to
`talk' to each other (finite $t^{\prime \prime}$, see green arrows in Fig. 1), i. e., if
an electron that tunnels into the CNT QD through one channel has a
finite probability of tunneling out through the other channel.
Figure 4 shows results comparing the SU(4) state ($t^{\prime}=0.2$,
$t^{\prime \prime}=0.0$) with the 2LSU(2) state ($t^{\prime}=t^{\prime \prime}=0.1$)
\cite{note2}. Notice that the conductances for the two states, although being both equal
to $2e^2/h$ at QF \cite{friedel},
are {\it qualitatively} very different: the 2LSU(2) state reaches a maximum of
$2e^2/h$, half of the maximum in the SU(4) state, and it resembles more the
results for a single-channel system \cite{discontinuity}. 

\begin{figure}[h]
\centering
\includegraphics[angle=90,height=5.0cm]{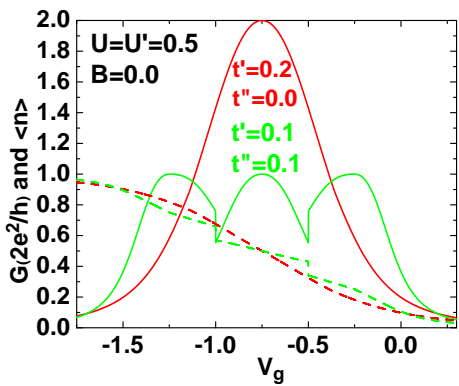} 
\caption{ Comparing SU(4) and 2LSU(2) conductances: The red
curves show the conductance $G$ (solid) and average occupancy $\left< n \right>$ (dashed)
for the SU(4) state (same as the red curves in Fig. 2), while the
green curves show results for 2LSU(2).
It is interesting to note that
for both states, when the CNT is occupied by one electron ($V_g=-0.25$
for 2LSU(2), and $V_g=-0.385$ for SU(4)), the conductance
reaches the unitary limit ($G=2e^2/h$), as discussed in ref. \onlinecite{aguado1},
however there is no question that the conductances for both states are qualitatively different.
}
\end{figure}

It is reasonable to assume that in a realistic experimental situation
the conservation of the orbital quantum number lies somewhere between the two schemes 
represented in fig. 1, therefore a careful analysis of the
robustness of the SU(4) state (in the presence of some channel 
mixing) is needed if one wants to correlate any of the
experimental observations to the results obtained by numerical modeling. Recent calculations
\cite{aguado1, aguado2} have analyzed the transition between these two states 
(SU(4) and 2LSU(2)) only at QF. In figure 5, we present 
complementary calculations for all fillings. Our results at QF confirm (as discussed below) that
SU(4) and 2LSU(2) at QF are experimentally indistinguishable, reinforcing then the need to extend
the analysis to other fillings, especially between QF and HF.

Figure 5 shows how the conductance evolves from the SU(4) to the
2LSU(2) state, for $t^{\prime \prime}$ varying from $0$ to $t^{\prime}$. Notice that,
as the value of $t^{\prime \prime}$ increases from zero (solid black curve),
the conductance peak at $4e^2/h$ (characteristic of the SU(4) state)
becomes gradually narrower, until (for $t^{\prime \prime}>0.175$) the
central peak splits into three very narrow peaks (not shown). An indication
of their presence can be seen already in the curve for
$t^{\prime \prime}=0.175$ (cyan), as indicated by the arrows.
As $t^{\prime \prime}$ approaches $t^{\prime}$, these three peaks continue to narrow, until they
vanish. On the other hand, for values of conductance around $2e^2/h$, the
curve develops shoulders which become broader as $t^{\prime \prime}$
increases. We want to stress the qualitative agreement of our QF results 
to those obtained by J.-S. Lim {\it et al.} \cite{aguado2}. 
In fig. 5, all curves cross at $V_g \approx -0.3$ (where $\langle n_d \rangle \approx 1$, QF) 
where they have approximately unitary conductance $G_0$. Therefore, as stressed in
references \onlinecite{aguado1} and \onlinecite{aguado2}, SU(4) and 2LSU(2) are experimentally
indistinguishable {\it at} QF and zero magnetic field.
In addition, it is also interesting to note, as described above, that our results for $t^{\prime \prime}
\approx t^{\prime}$ change discontinuously to the 2LSU(2) result (dashed curve). 
A similar discontinuity is seen in the Slave Boson Mean Field results at QF in reference
\onlinecite{aguado2} (please, check their fig. 14).

Finally, the green solid curve in fig. 4 has a discontinuity in the
conductance for $V_g = -0.5$. We have seen this kind 
of behavior in other multi-orbital systems \cite{discontinuity}, 
and in some cases associated it to the crossing through
the Fermi energy of a very narrow level. This causes an
abrupt charging of the QD (clearly visible
in the green dashed curve for $\langle n_d \rangle$ vs. $V_g$), 
with the consequent abrupt change of the conductance. In this specific case, this level 
can be identified to the $\epsilon_-$ energy level in the upper
panel of fig. 11 in reference \onlinecite{aguado2}, which, 
as described there, has a vanishingly narrow width.

\begin{figure}[h]
\centering
\includegraphics[angle=90,height=5.0cm]{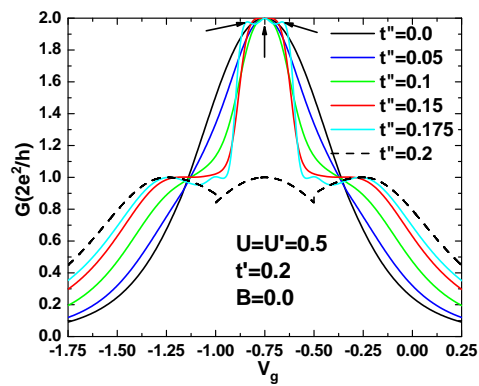}
\caption{ Transition from the SU(4) to the 2LSU(2) state:
Results showing the variation of the conductance for $t^{\prime \prime}$
varying from 0.0 (black curve) to 0.2 (dashed), going through the values
0.05 (blue), 0.1 (green), 0.15 (red) and 0.175 (cyan). $U = U^{\prime} = 0.5$ and
$t^{\prime}=0.2$.
}
\end{figure}

\section{Experimental observation of SU(4) Kondo at Half-Filling}

Figure 6a reproduces Fig. SI2 in the Supplementary Information of 
ref. \onlinecite{jarillo}. In it, the temperature variation of the 
conductance for three shells in the CNT QD is reported (thick red line 
at higher temperature and black thin line for the lowest temperature). 
Notice that the coupling to the leads increases from right to left \cite{jarillo,jarillo2} 
(that is why the conductance of shell 1 is the lowest). In reference 
\onlinecite{jarillo}, the conductance in regions I and III in the second 
shell was associated to a QF SU(4) state and the conductance at HF in 
shell $n=3$ was associated to a singlet-triplet effect \cite{sasaki,yeyati}. However, 
in Fig. 6b, our results indicate an alternative interpretation: by breaking 
the degeneracy of the orbital levels (by introducing a small energy 
splitting $\delta E = 0.032$) \cite{splitting} and by increasing the coupling of the 
third shell to the leads ($t^{\prime}=0.2$) in relation to the second 
shell ($t^{\prime}=0.11$), the experimental results can be qualitatively reproduced. 
Note that since our calculations are done at zero-temperature, the curve in Fig. 6b should 
be compared to the highest conductance curve in Fig. 6a (thin black line).
It is clear that the qualitative agreement is quite good.

\begin{figure}[h]
\centering
\includegraphics[height=3.3cm]{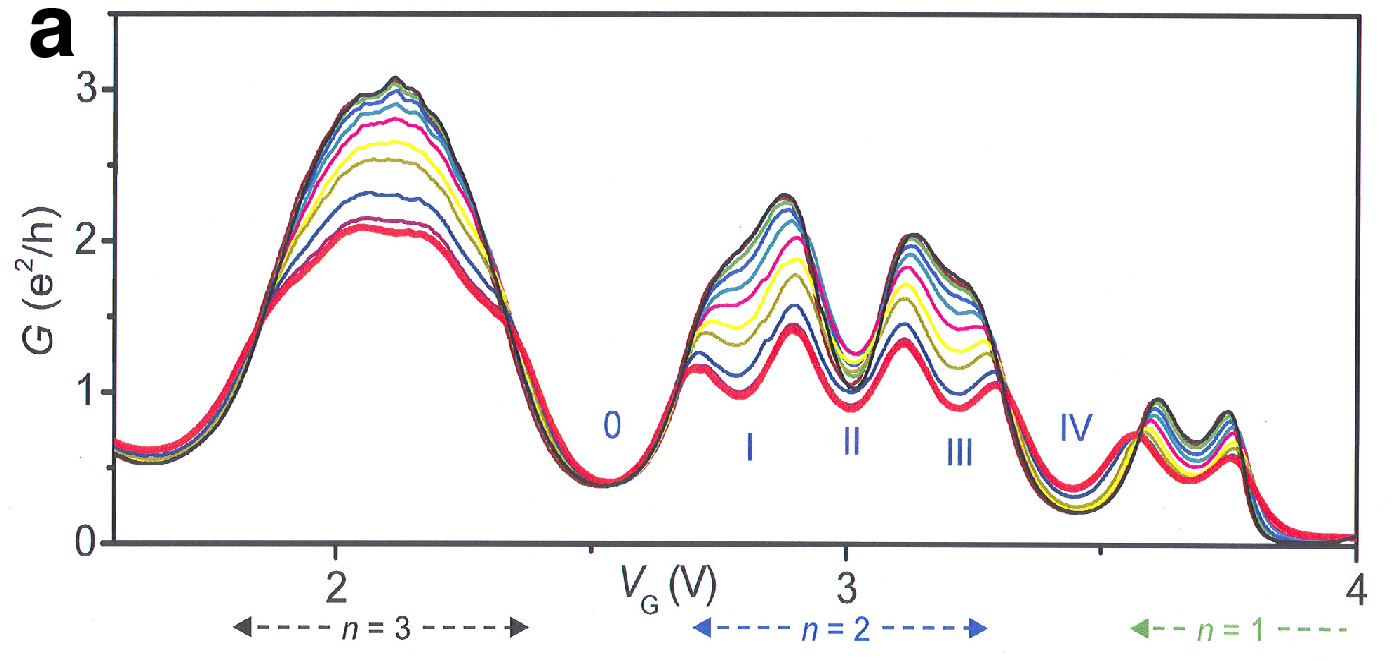}
\includegraphics[angle=90,height=4.5cm]{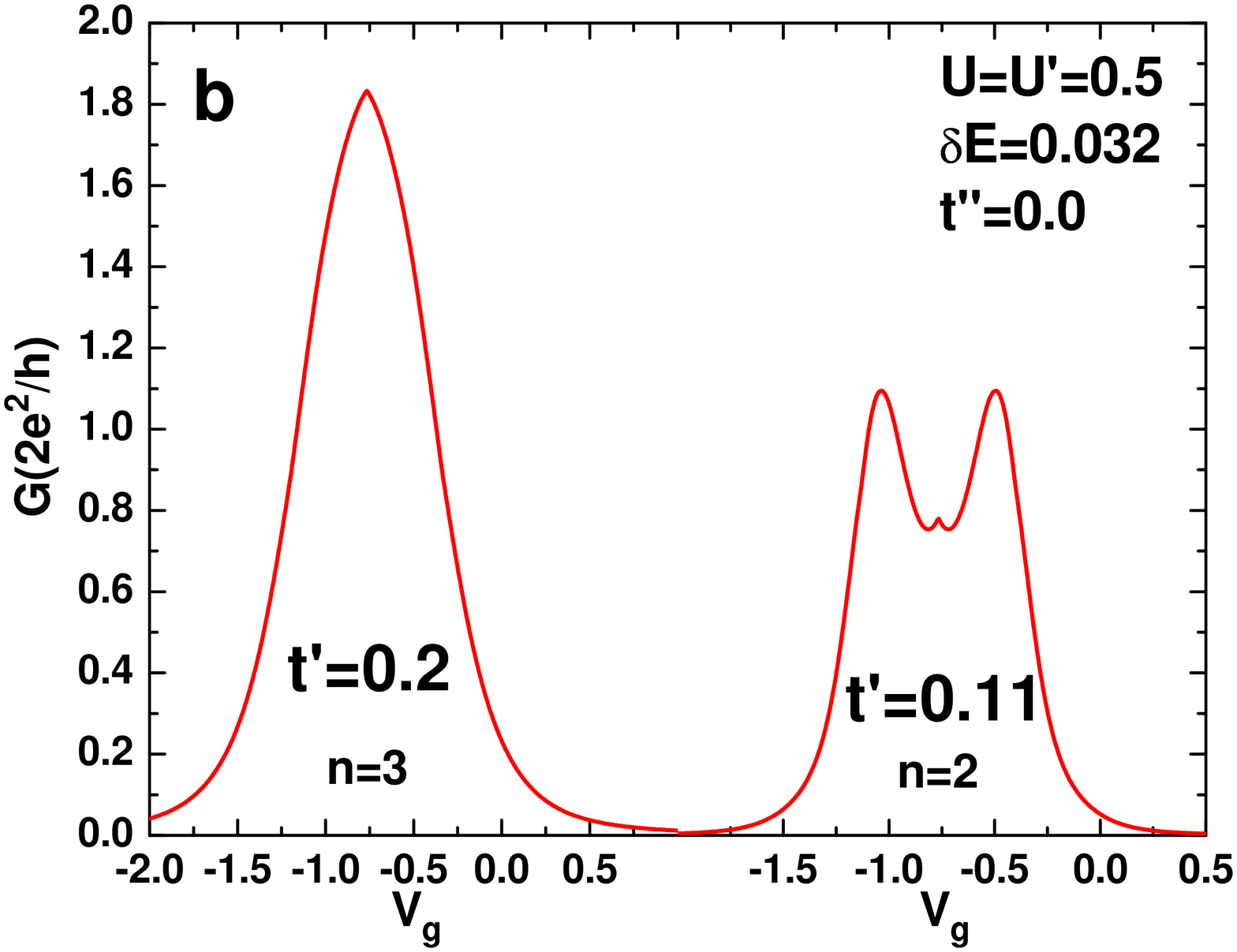}
\caption{ (a)  Adapted from Supplementary Information in ref. \onlinecite{jarillo} (Fig. SI2):
temperature dependence of the conductance for the three shells observed in
the CNT QD. Thick red curve at higher temperature and thin black curve at lowest temperature.
Note that the coupling to the leads increases from right to left (from shell
$n=1$ to $n=3$) \cite{jarillo,jarillo2}.
In ref. \onlinecite{jarillo}, the regions indicated as I and III in the second shell were
interpreted as indicative of the presence of a QF SU(4) state and
the conductance of the third shell (leftmost) was associated to
a singlet-triplet  state \cite{sasaki}, however,
our simulations in (b) show that the conductance for both shells can be interpreted
as indication of a QF SU(4) state for the second shell and
of {\it both} QF and HF SU(4) states for the third shell. These results are obtained
by adding a small energy separation between the orbital levels
($\delta E=0.032$) and by increasing the coupling of the third shell
($t^{\prime}=0.2$) in relation to the second shell ($t^{\prime}=0.11$).
}
\end{figure}

To further test our simulations against the experimental results, 
Fig. 7a shows a color-scale plot with numerical results for
the variation of the conductance (at zero bias) with applied magnetic field 
(along the CNT axis) for the third shell (same parameters as the ones used 
in Fig. 6b). 
Since the orbital moment is much larger than
the spin one ($\mu_{orb}=0.2$ and $\mu_{sp}=0.04$), at lower fields
one sees first the splitting of the SU(4) conductance peak into two spin SU(2) Kondo
peaks, which at higher field values will each further split into two CB peaks.
Figure 7b shows a figure adapted from ref. \onlinecite{jarillo2} containing
field-dependent conductance results (third shell in Fig. 6a)
which are clearly in qualitative agreement with the numerical results in Fig. 7a.
The combined results presented in Figs. 6 and 7 present compelling evidence that 
the conductance results of the third shell can be interpreted as a manifestation 
of the HF SU(4) state. It is interesting to note the asymmetry in the 
conductance in the experimental results in Fig. 7b, i.e., the CB regime 
is reached at lower values of field for lower values of gate voltage. 
The same kind of asymmetry was observed by Makarovski {\it et al.} \cite{gleb} 
in all shells and always with the higher conductance for 3 electrons in the shell. 
It is not clear yet the reason for this higher Kondo temperature for 3 electrons 
when compared to QF (1 electron inside the shell). 

\begin{figure}[h]
\centering
\includegraphics[angle=90,height=3.5cm]{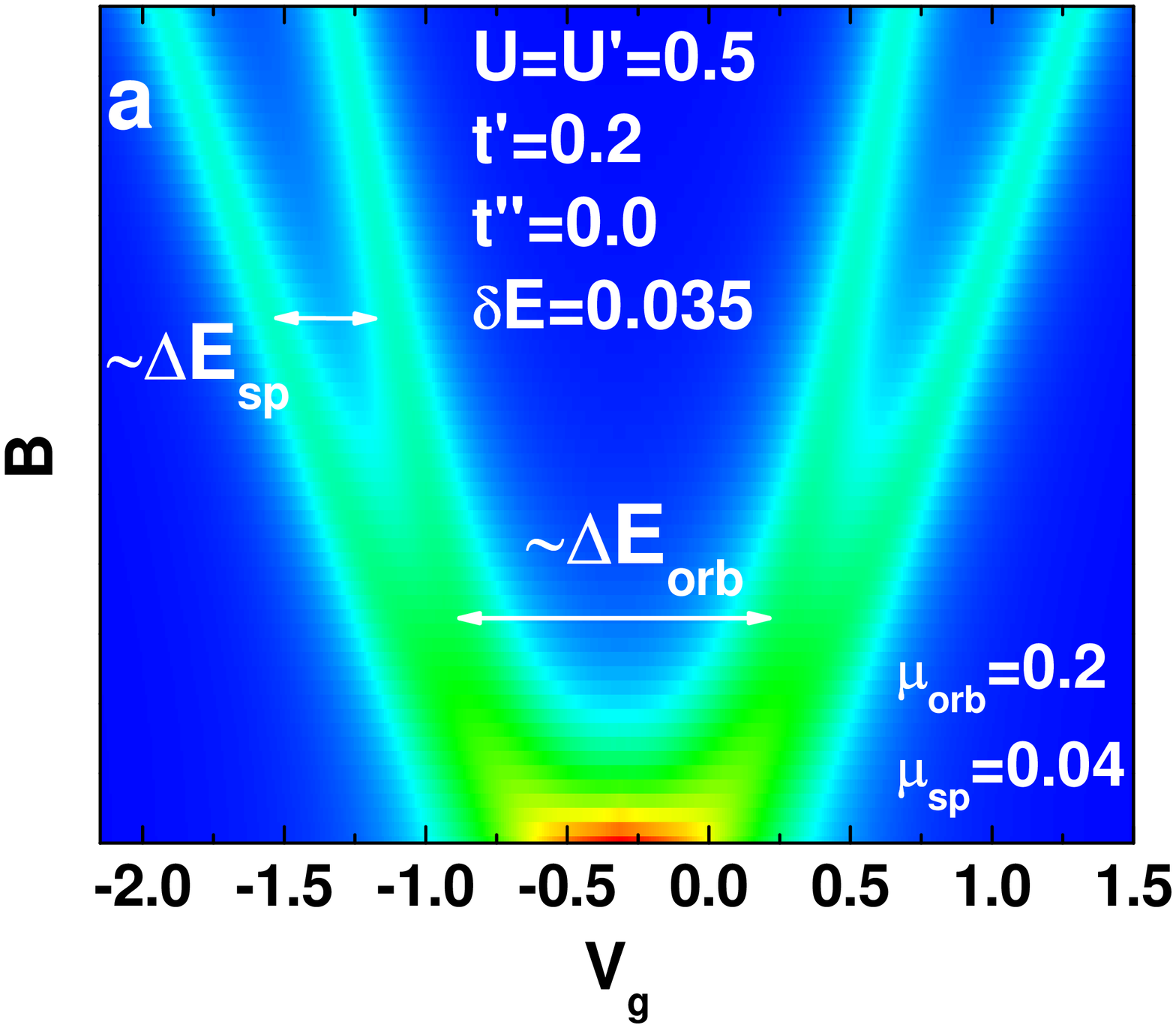}
\includegraphics[height=2.5cm]{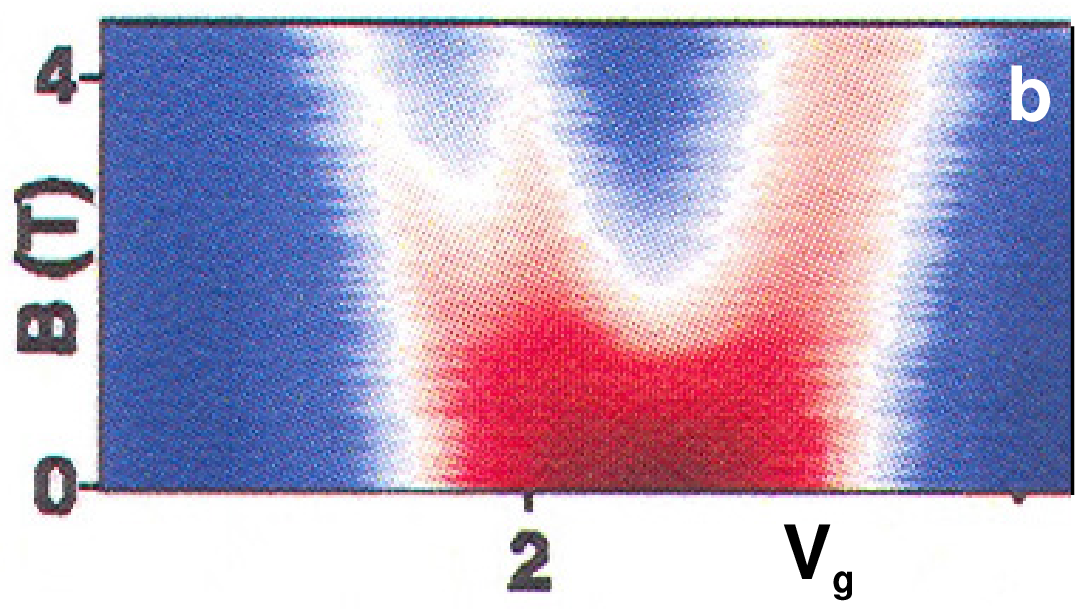}
\caption{ Splitting of the Kondo SU(4) peak caused by an external magnetic 
field applied along the CNT axis. (a) Color-scale plot of the conductance 
showing the progressive splitting of the zero-field SU(4) peak into 
two spin SU(2) peaks and then into 4 CB peaks 
(dark red $\approx 4e^2/h$ and blue $=0$). (b) Experimental results 
adapted from ref. \onlinecite{jarillo2} (same shell as the leftmost one 
in Fig. 6a) indicating that the numerical results in part (a) qualitatively 
reproduce the experiments.
}
\end{figure}

\section{Conclusions}

In summary, using a recently developed numerical method (ECA) \cite{method}, 
the authors offer a reinterpretation of recent 
transport measurements in CNT QDs. In these experiments, 
the conductance of one of the electronic shells of a CNT QD was 
associated to the SU(4) Kondo effect at QF, and  
the conductance at HF of an adjacent shell was interpreted as 
resulting from a singlet-triplet effect \cite{jarillo}. 
Our results clearly show that the conductance of {\it both} 
shells can be interpreted instead as resulting from an SU(4) state\cite{note-st}: 
this is achieved by introducing a small energy splitting between the orbital levels 
and increasing the coupling to the leads of one of the shells in relation to the other 
(see Fig. 6).  
Furthermore, simulations of conductance at finite magnetic field give support 
to our interpretation. These results open the possibility that the SU(4) state at 
HF could be analyzed in detail in CNT QDs. The fact that recent NRG results 
by Galpin {\it et al.} \cite{galpin1} 
have associated this state to rich physical behavior, including a non-Fermi-liquid phase, adds 
to the importance of our results. In addition, simulations presented in 
Fig. 2 are in qualitative agreement with Galpin {\it et al.}'s NRG results. 
This suggests that the low-energy physics associated to their NRG results is quite 
robust and should in principle be observed experimentally.

One last point the authors would like to stress is related to the implications 
of the SU(4) `spin-orbital entanglement' to the structure of the Kondo cloud.
A qualitative description of the screening mechanism in the Kondo effect involves the existence 
of the so-called `Kondo cloud': associated to the Kondo temperature $T_K$ 
(a universal energy scale which emerges naturally from Renormalization Group arguments) 
there is a universal length scale $\xi_K = \hbar v_F/T_K$, which can be interpreted 
as the size of the many-body wave function containing the conduction electron 
that forms a singlet with the impurity spin. Even before the first measurement of the Kondo effect in 
semiconducting QDs \cite{goldhaber2}, there was great interest in experimentally detecting the Kondo cloud, 
with theoretical \cite{bergmann} and experimental \cite{giordano} 
efforts being made to evaluate and measure its size and dependence on dimensionality. 
The failure to actually observe the extent of the Kondo cloud (or even ascertain its existence) 
underscores the difficulties involved.
This has lead more recently to theoretical efforts to analyze setups where the Fermi 
sea is effectively `confined' {\it inside} the nanostructure, like for example 
in the so-called `Kondo Box' \cite{thimm} or in QDs embedded in Aharonov-Bohm (AB) rings \cite{affleck}. 
In such setups, interesting new effects are expected to occur, hopefully leading to a 
better understanding of the screening effect and a possible direct 
or indirect breakthrough measurement of the Kondo cloud. The difficulty of fabricating 
the proposed devices and performing the necessary measurements may explain the fact that most of 
the work in this area is theoretical. 
Referring back to fig. 1, the many-body wave function formed by conduction electrons 
that screen the localized moment has to have quite different properties 
in the SU(4) Kondo state when compared to the 2LSU(2) state. 
Indeed, the recognition that orbital Kondo correlations can only form if the orbital QN 
is conserved upon tunneling, and that electron states in the metallic leads do 
not have a defined `wrapping mode', results in the natural conclusion that in the SU(4) Kondo 
state the `Fermi sea' is in reality formed by 
electrons that have a well defined orbital QN, and therefore should reside 
primarily in the regions of the CNT contained between the tunnel
barrier and the metallic contacts \cite{jarillo, aguado1,aguado2}. These regions, as 
for example in the proposed setups involving QDs embedded in AB rings \cite{affleck}, 
naturally constrain the extent of the Kondo cloud. Obviously, for CNT devices where 
the tunneling barrier is exactly at the interface between the CNT and the metallic 
contacts, there is no conservation of the orbital QN, leading to
a 2LSU(2) state, where there is no entanglement of
the spin and orbital degrees of freedom and therefore its associated Kondo cloud
is free to spread inside the metallic contacts. 
One of the problems with the experimental realization of setups suggested as 
possible probes of the Kondo cloud properties is that, once leads are attached 
to the nanostructure to perform actual measurements, the Kondo cloud spreads into them, 
making the measurements of its properties difficult. 
What we suggest here is that the SU(4) state in carbon nanotubes naturally provides 
a system where the Kondo cloud should be constrained inside the nanostructure 
itself, with the advantage that the leads needed for conductance 
measurements, at least in principle, do not `accept' the Kondo cloud. 
In that case, careful analysis of the change in transport properties as the system 
transitions from the SU(4) to the 2LSU(2) state should provide valuable information about the screening 
process and at least some indirect information about the Kondo cloud.

The authors acknowledge useful discussions with E. Anda, E. Dagotto, G. Finkelstein, 
D. Goldhaber-Gordon, P. Jarillo-Herrero and J. Riera.
G. B. M. acknowledges support from Research Corporation; 
C. A. B. acknowledges support from UT, Knoxville.


\begin{thebibliography}{}
\bibitem{note1} The Kondo temperature observed in the orbitally degenerate CNT QD 
was as high as $T_{\rm K}=16.0 {\rm K}$ \cite{jarillo}.
\bibitem{range} As noted in reference \onlinecite{galpin1}, one expects that the 
SU(4) state on each side of the $U=U^{\prime}$ point will survive for an interval 
$| U^{\prime}-U | \approx T_{K}^{SU(4)}$.
\bibitem{note2} A smaller value is used for the hoppings in the 2LSU(2) state 
to compensate for the fact that the number of hoppings between the QD and
the leads in 2LSU(2) is double that in SU(4) (see Fig. 1).
\bibitem{friedel} This can be obtained in a very simple way by invoking the
Friedel sum rule (check reference \onlinecite{aguado1}).
\bibitem{note-st} This claim can be tested by measuring the variation of $T_K$
away from the orbital degeneracy point: it should be asymmetric for 
ST and symmetric for SU(4), as indicated in reference \onlinecite{sasaki1}.
\end{thebibliography}
\end{document}